\documentclass[aps,pra,twocolumn]{revtex4}
\usepackage{graphicx}
\usepackage{bm}

\usepackage{caption}
\usepackage{subcaption}
\usepackage{amsmath,amssymb,amsthm,dsfont,bm}
\usepackage{color}
\usepackage{wasysym}
\usepackage{ulem}
\usepackage{verbatim}
\usepackage{soul}

\usepackage[applemac]{inputenc}
\usepackage[dvipsnames]{xcolor}
\usepackage[colorlinks=true,urlcolor=blue,citecolor=blue,linkcolor=blue]{hyperref}

\begin{document}

\title{Beam splitter as quantum coherence-maker}
\author{Laura Ares}\email{laurares@ucm.es} 
\author{Alfredo Luis}\email{alluis@ucm.es}
\affiliation{Departamento de \'Optica, Facultad de Ciencias
F\'{\i}sicas, Universidad Complutense, 28040 Madrid, Spain}
\date{\today}

\begin{abstract}
The aim of this work is to answer the question of how much quantum coherence a beam splitter is able to produce. To this end we consider as the variables under study both the amount of coherence of the input states as well as the beam splitter characteristics. We conclude that there is an optimal combination of these factors making the gain of coherence maximum. In addition, the two mode squeezed vacuum arises as the studied state most capable of gaining coherence when passing through a beam splitter. These results are qualitatively equivalent for the the l1-norm of coherence and for the relative entropy of coherence. 

\end{abstract}

\maketitle

\section{Introduction}

Quantum coherence is the archetype of quantum correlations. It potentially concentrates all the superposition principle consequences, meaning all the fascinating fundamental implications \cite{AK21, XY19, MS04} along with the prospering practical applications \cite{FC21, TM19, CL21}. This general role allows coherence to be transformed into other types of correlations such as entanglement or steering \cite{KXK21,WWM16,JM16}.

Quantum resources theories \cite{SP17,CG19,Tan19} have proved to be a useful support to take advantage of all these quantum features, for example in metrology \cite{WG20}  or quantum information processing \cite{YQN19}. The general scheme of these theories is common, broadly, to quantify the resource by defining the states lacking the resource, the operations that do not generate it, and finally the quantifier that accounts for the amount of resource present \cite{HH09,BCP14}. In this work we focus on characterize one operation that is not free with regard to quantum coherence, the beam splitter, which is able to enhance the coherence previously present in the system \cite{ZSLF16, AL22}. This behaviour of beam splitter as coherence-maker is already present in classical optics. Moreover, beam splitters are the key element involved in generating other quantum correlations from quantum coherence \cite{GH21,WWM16}. These two properties make us wonder how the beam splitter parameters and the input states influence the increase in coherence.

Accordingly, we investigate which choice of the reflectance-transmittance ratio and phase changes produce maximum coherence for a given input state. As the input states we consider incoherent as well as coherent states, so we can evaluate the final amount of coherence regarding the initial one. We also divide the problem into subspaces of fixed incoming energy and compare the possible distributions of photons between the two input modes.
Since coherence is a basis dependent quantity, throughout this analysis we are working in the photon number basis so we are always talking about photon-number-coherence. The quantifiers of coherence utilized are the l1-norm and the relative entropy of coherence \cite{BCP14}.

\section{Beam splitter}

\begin{figure}[h]
    \includegraphics[width=6cm]{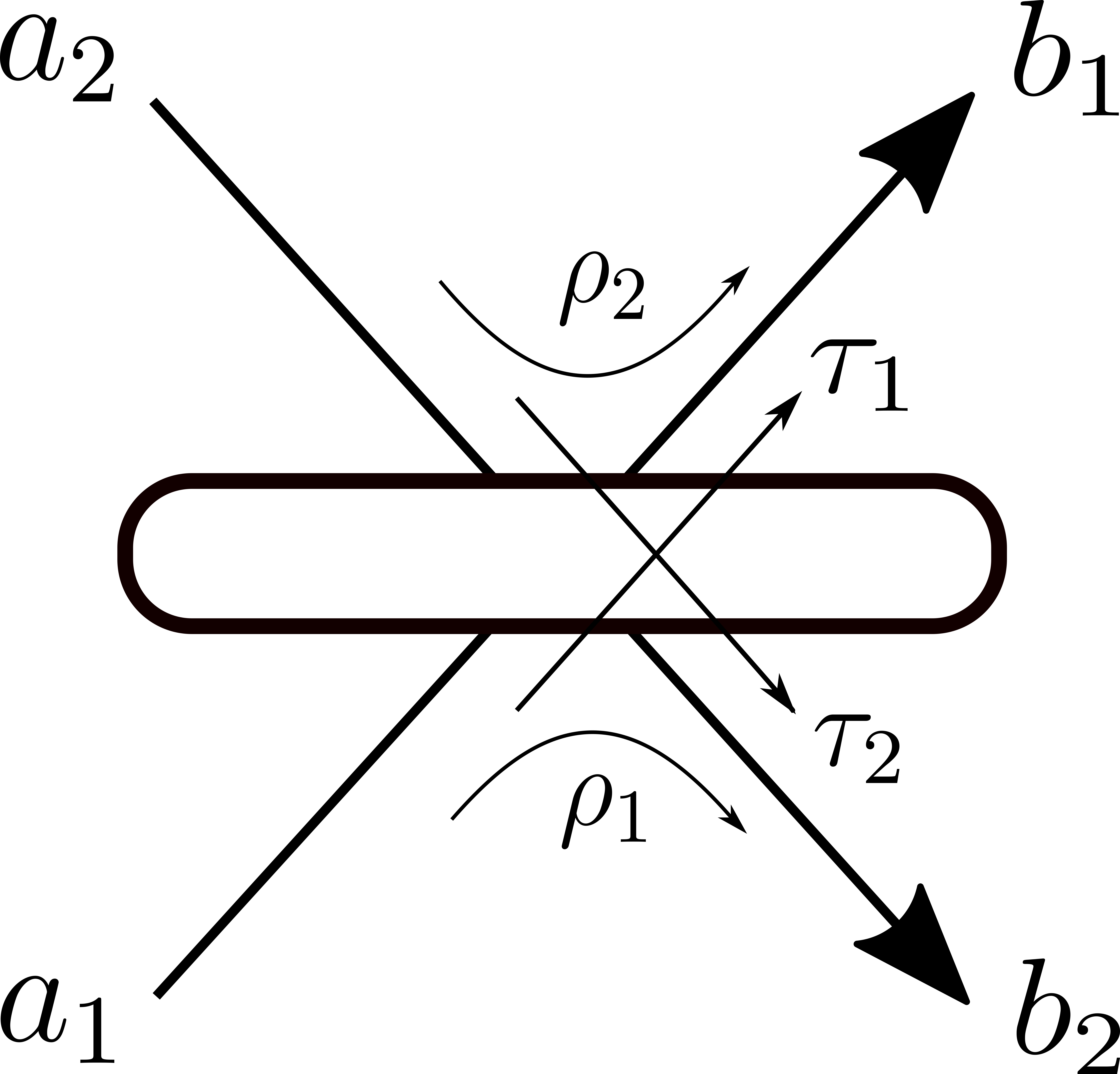}
    \caption{Scheme of a beam splitter specifying the modes involved and the definition of the transmission and reflection coefficients $\tau_{1,2}$ $\rho_{1,2}$.  }
    \label{bs}
\end{figure}

The action of the beam splitter can be expressed in matrix form as a relation between the input $a_{1,2}$ and output $b_{1,2}$ complex-amplitude operators 
\begin{equation}
\label{io}
 \begin{pmatrix} b_1\\b_2 \end{pmatrix} = \begin{pmatrix} \tau_1 & \rho_2 \\ \rho_1 & \tau_2 \end{pmatrix}  \begin{pmatrix}a_1 \\ a_2 \end{pmatrix} ,
\end{equation}
where $\tau_{1,2}$ $\rho_{1,2}$ are the corresponding complex transmission and reflection coefficients as defined in Fig. \ref{bs}. For a lossless beam splitter the transformation matrix is unitary, which means \cite{AS95}
\begin{equation}
\label{ph}
    |\tau_1|= |\tau_2 | = \cos \theta , \quad |\rho_1|= |\rho_2 | = \sin \theta ,
\end{equation}
where we have introduced the parameter $\theta$ to express the balance between transmission and reflection, with $\theta \in [0,\pi/2]$, and 
\begin{equation}
\label{pr}
    \delta_{\rho_1}- \delta_{\tau_2}+\delta_{\rho_2}-\delta_{\tau_1} = \pi ,
\end{equation}
where $\delta$ are the phases of the corresponding coefficients. 

{}
\section{Incoherent input state}

We start this analysis by considering a beam splitter illuminated by a number state on both input ports.

The corresponding incoming  number state  $|n_1\rangle_1\otimes|n_2 \rangle_2=|n_1,n_2 \rangle$ can be expressed in terms of the vacuum as 
\begin{equation}
\label{in} 
  |n_1,n_2 \rangle = \frac{1}{\sqrt{n_1! n_2!}} a_1^{\dagger n_1} a_2^{\dagger n_2} |0,0\rangle .
\end{equation}

The output state can be readily obtained in the number basis by inverting the input-output relation (\ref{io}) to express the input modes $a_{1,2}$ in terms of the output modes $b_{1,2}$, and translating the result to Eq. (\ref{in}) to get

\begin{widetext}

\begin{equation}
\label{out}
  | {\rm out} \rangle_{n_1,n_2} = \frac{1}{\sqrt{n_1! n_2!}} \left ( e^{i\delta_{\tau_1}} \cos \theta b_1^\dagger + e^{i \delta_{\rho_2}} \sin \theta b^\dagger_2  \right )^{n_1} \left ( e^{i \delta_{\rho_2}} \sin \theta b_1^\dagger + e^{i \delta_{\tau_2}} \cos \theta b^\dagger_2  \right )^{n_2}|0,0\rangle .
\end{equation}
Equivalently, the decomposition formulas for the su(2) Lie algebras may be used \cite{MB93} . After some simple algebra, and using relation (\ref{pr}) we get that the output state (\ref{out}) becomes 
\begin{equation}
  | {\rm out} \rangle_{n_1,n_2} = \sum_{j=0}^{n_1+n_2} c_j  | j, n_1+n_2 - j \rangle ,
\end{equation}
where 
\begin{equation}
\label{cj}
 c_j = \sqrt{n_1! n_2!j!(n_1+n_2-j)!} e^{ij(\delta_{\tau_1} - \delta_{\rho_2})} \sum_{k={\rm max} (0,j-n_2)}^{n_1} \frac{(-1)^k \cos^{n_2+2k-j} \theta \sin^{n_1-2k+j} \theta }{(n_1-k)! k! (n_2+k-j)!(j-k)!} .
\end{equation}

\end{widetext}

In the particular case of $n_1=0$ or $n_2= 0$ we get that $| {\rm out} \rangle$ are SU(2) coherent states \cite{ACGT72}. The computation of the coherence depends just on the modulus of $|c_j|$ so that the actual value of the phase $\delta_{\tau_1} - \delta_{\rho_2}$ is irrelevant. 

\bigskip

As suitable coherence measures we have the $l_1$-norm of coherence, $C_H$,  and the relative entropy of coherence, $C_S$, which we will use in their forms adapted to pure states \cite{BCP14,SP17} 
\begin{equation}
    \mathcal{C}_H = \left ( \sum_j |c_j |\right )^2 -1,
\end{equation}
and
\begin{equation}
      \mathcal{C}_S = - \sum_j |c_j|^2 \ln |c_j|^2 ,
\end{equation}
respectively. If we restrict ourselves to a subspace of fixed energy, $n_1+n_2={\rm constant}$, the maximum coherence holds for $|c_j | = \mathrm{constant}$ \cite{CH15,AS18}, this is for the phase-like states 
\begin{equation}
\label{pls}
    |\bm{\phi} \rangle = \frac{1}{\sqrt{n_1+n_2+1}} \sum_{j=0}^{n_1+n_2} e^{i \phi_j}  | j, n_1+n_2 - j \rangle,
\end{equation}
where $\phi_j$ are arbitrary phases \cite{FDPS,FDPS2}. The corresponding maximum of coherence is 
\begin{equation}
\label{mc}
     \mathcal{C}^{\rm max}_H = n_1 + n_2 ,
\end{equation}
leading to a curious but accidental identification of coherence with energy. In the case of the relative entropy of coherence the maximum value is
\begin{equation}
\label{CSmax}
    C_S^{\rm max}=-{\ln}\left[\frac{1}{n_1+n_2+1}\right].
\end{equation}

As we are about to see, not all incoming states are able to reach this maximum coherence, irrespective of the beam splitter parameters.
 This idea leads us to an alternative expression for the coherence as a the maximum overlap between the system state $|\psi \rangle$, assumed pure, and the phase-like states $|\bm{\phi} \rangle$ when $\bm{\phi}$ is varied 
\begin{equation}
\label{overlap}
    \mathcal{C}_H =(\sqrt{n_1+n_2+1}\hspace{1mm} \mathrm{ max}_{\bm{\phi}} \langle \bm{\phi} |\psi \rangle )^2-1.
\end{equation}

\subsection{One photon}
This case, $n_1+n_2 =1$, is a rather simple situation since essentially there is just a single configuration, the input number state $|1,0\rangle$. The corresponding output state, omitting irrelevant relative phases, is the split photon 
\begin{equation}
  | {\rm out} \rangle_{1,0} = \cos \theta |1, 0 \rangle + \sin \theta |0,1 \rangle ,
\end{equation}
that without any calculus shows that maximum coherence holds for a 50 \% beam splitter $\theta= \pi/4$, for which $| {\rm out} \rangle_{1,0}$ becomes a phase-like state $ |\bm{\phi}  \rangle$ in Eq. (\ref{pls}).

\subsection{Two photons}
In this situation where $n_1+n_2 =2$ we have just two meaningful cases, say SU(2) coherent states  $|2,0 \rangle$ and SU(2) squeezed states $|1,1 \rangle$  \cite{ACGT72,AP94,BM96}, with output states 
\begin{eqnarray}
  &| {\rm out} \rangle_{2,0} = \\
  &\cos^2 \theta |2, 0 \rangle + \sqrt{2} \sin \theta \cos \theta  |1,1 \rangle + \sin^2 \theta |0,2 \rangle \nonumber ,
\end{eqnarray}
and 
\begin{eqnarray}
  & | {\rm out} \rangle_{1,1} = \sqrt{2}  \sin \theta \cos \theta  |2, 0 \rangle +    \\
 &  \left (  \cos^2 \theta - \sin^2 \theta  \right ) |1,1 \rangle- \sqrt{2}  \sin \theta \cos \theta  |0,2 \rangle .  \nonumber 
\end{eqnarray}

\begin{figure}[h]
        \includegraphics[width=8cm]{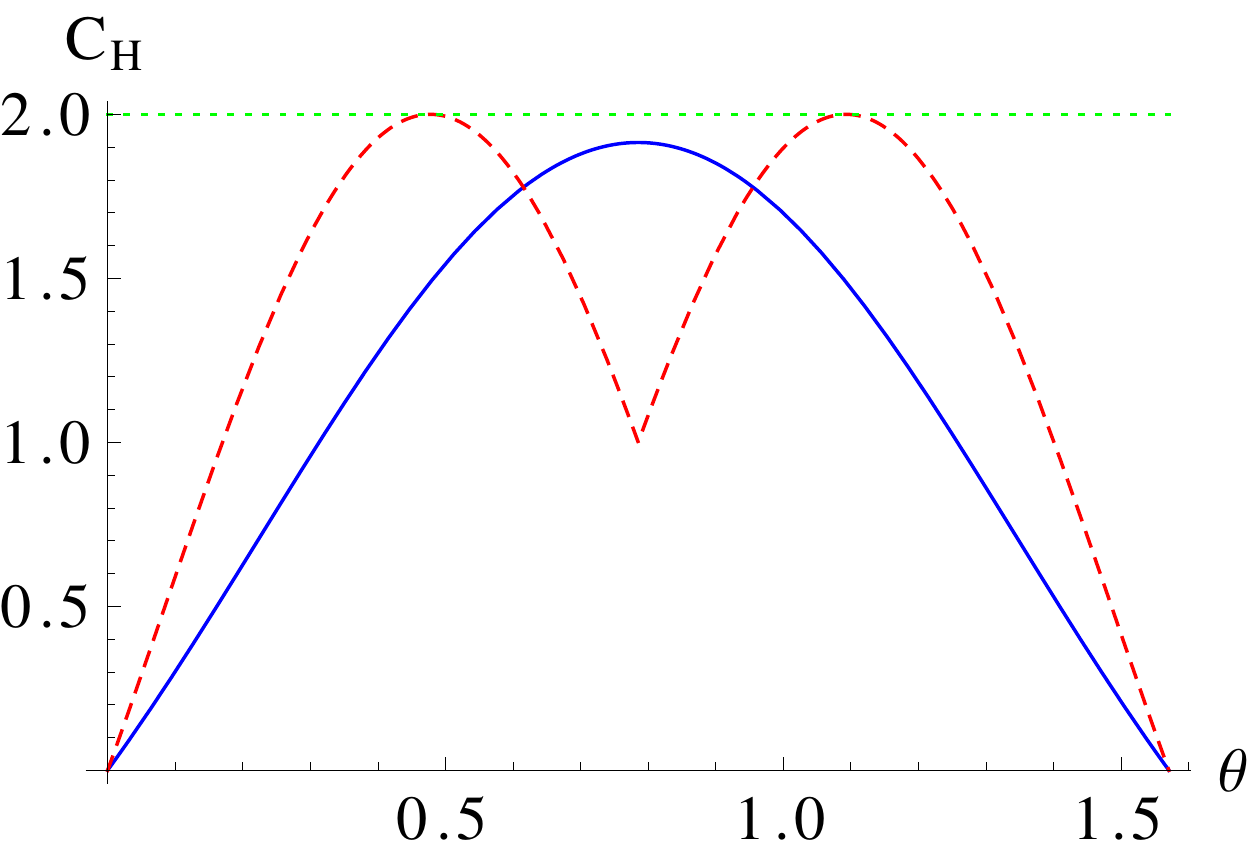}
        \caption{ $\mathcal{C}_H$  for the states  $ | {\rm out} \rangle_{2,0}$ (blue solid line) and  $ | {\rm out} \rangle_{1,1}$ (red, dashed line) as functions of the beam splitter's parameter $\theta$. The green, dotted line marks the maximum coherence (\ref{mc}).}
        \label{CH2}
    \end{figure}

In Fig. \ref{CH2} we represent the coherence $\mathcal{C}_H$ for these states as a function of $\theta$, blue solid line for the $ | {\rm out} \rangle_{2,0}$ and red dashed line for $| {\rm out} \rangle_{1,1}$, while the green line marks the maximum coherence in Eq. (\ref{mc}).

We can appreciate that varying the transmittance-reflectance ratio of the beam splitter, the $| {\rm out} \rangle_{1,1}$ state can get larger coherence than the SU(2) coherent state $ | {\rm out} \rangle_{2,0}$, which betrays its name a bit.

\bigskip

Moreover, it can be clearly seen that the SU(2)-coherent case $ | {\rm out} \rangle_{2,0}$ finds its maximum coherence for a balanced beam splitter $\theta = \pi/4$. On the other hand, the coherence for the  $|{\rm out} \rangle_{1,1}$ case finds its maximum for an unbalanced disposition, that can be found analytically to be given by the $\theta$ satisfying the equality $\tan ( 2 \theta) = \pm \sqrt{2}$. For these optimum beam splitters we have that the output are actually phase-like states (\ref{pls})
\begin{equation}
  | {\rm out} \rangle_{1,1} = \frac{1}{\sqrt{3}} \left ( \pm |2, 0 \rangle + |1,1 \rangle \mp  |0,2 \rangle \right ),
\end{equation}
while for a balanced beam splitter $\theta=\pi/4$ we will have the N00N state 
\begin{equation}
  | {\rm out} \rangle_{1,1} = \frac{1}{\sqrt{2}} \left ( |2, 0 \rangle-  |0,2 \rangle \right ).
\end{equation}

\begin{figure}
    \centering
    \includegraphics[width=8cm]{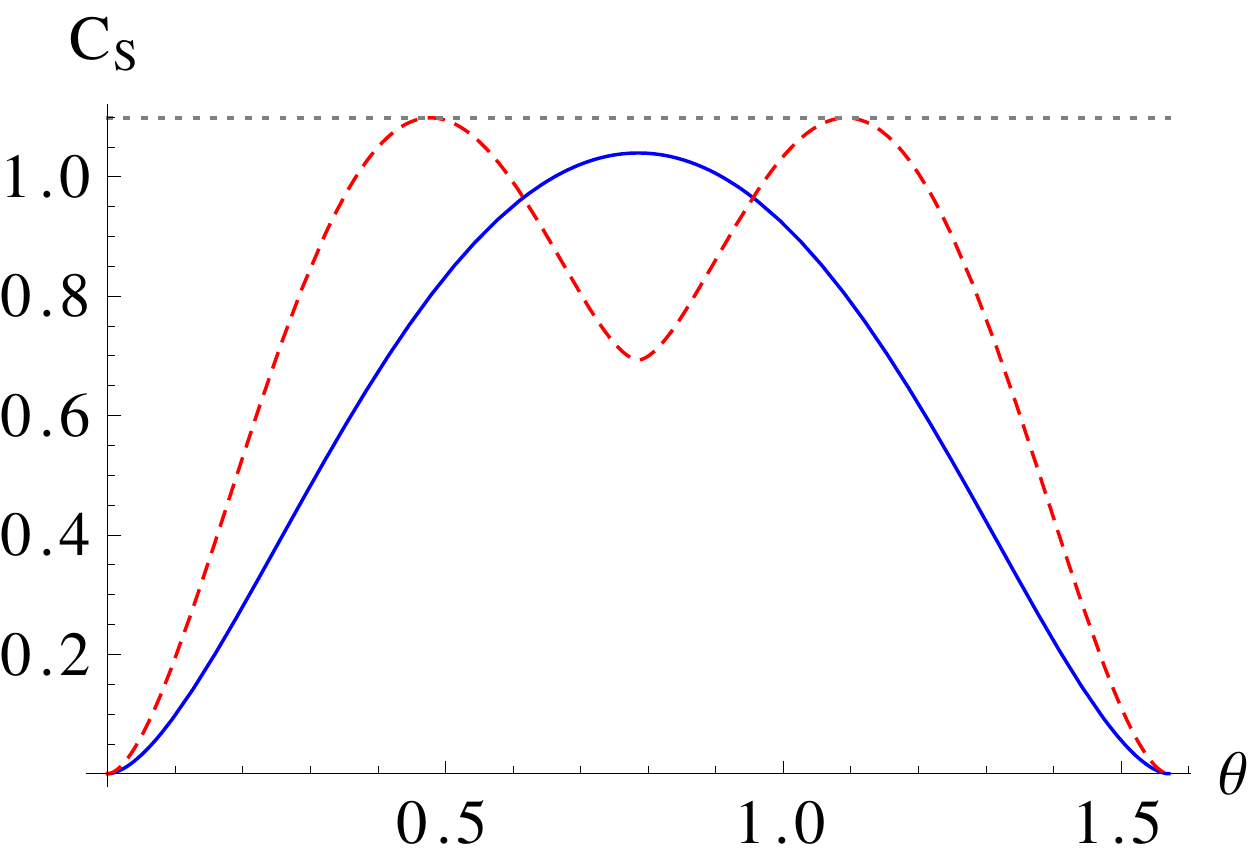}
    \caption{ $\mathcal{C}_S$  for the states  $ | {\rm out} \rangle_{2,0}$ (blue solid line) and  $ | {\rm out} \rangle_{1,1}$ (red, dashed line) as functions of the beam splitter's parameter $\theta$. The gray, dotted line marks the maximum coherence achieved.}
    \label{CS2}
\end{figure}

This result is reproduced if we consider the relative entropy of coherence. As expected, the absolute values of $C_H$ and $C_S$ are different, however the tendencies and the optimum configurations of the beam splitter are alike. From  Fig. \ref{CS2} we can highlight that the behaviour around the balanced beam splitter is much smoother in this case. 

\subsection{Larger photon numbers }

\begin{figure*}
    \begin{subfigure}[t]{0.32\textwidth}
        \includegraphics[width=\textwidth]{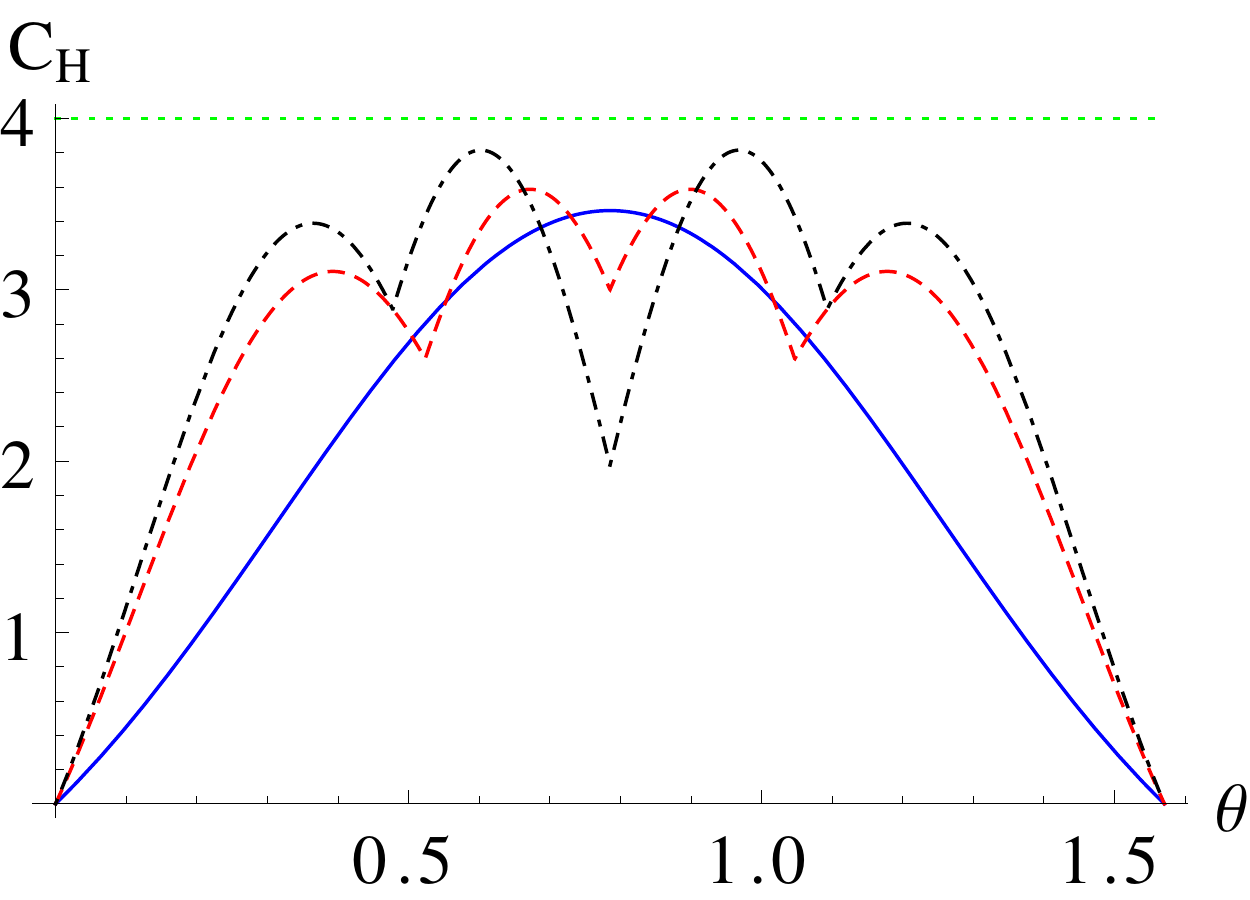}
        \caption{ $C_H$ for the states  $ | {\rm out} \rangle_{4,0}$ (blue solid line), $| {\rm out} \rangle_{3,1}$ (black, dotted-dashed line), and $ | {\rm out} \rangle_{2,2}$ (red,dashed line)}
        \label{CH4}
    \end{subfigure}{}\hfill
    \begin{subfigure}[t]{0.32\textwidth}
        \includegraphics[width=\textwidth]{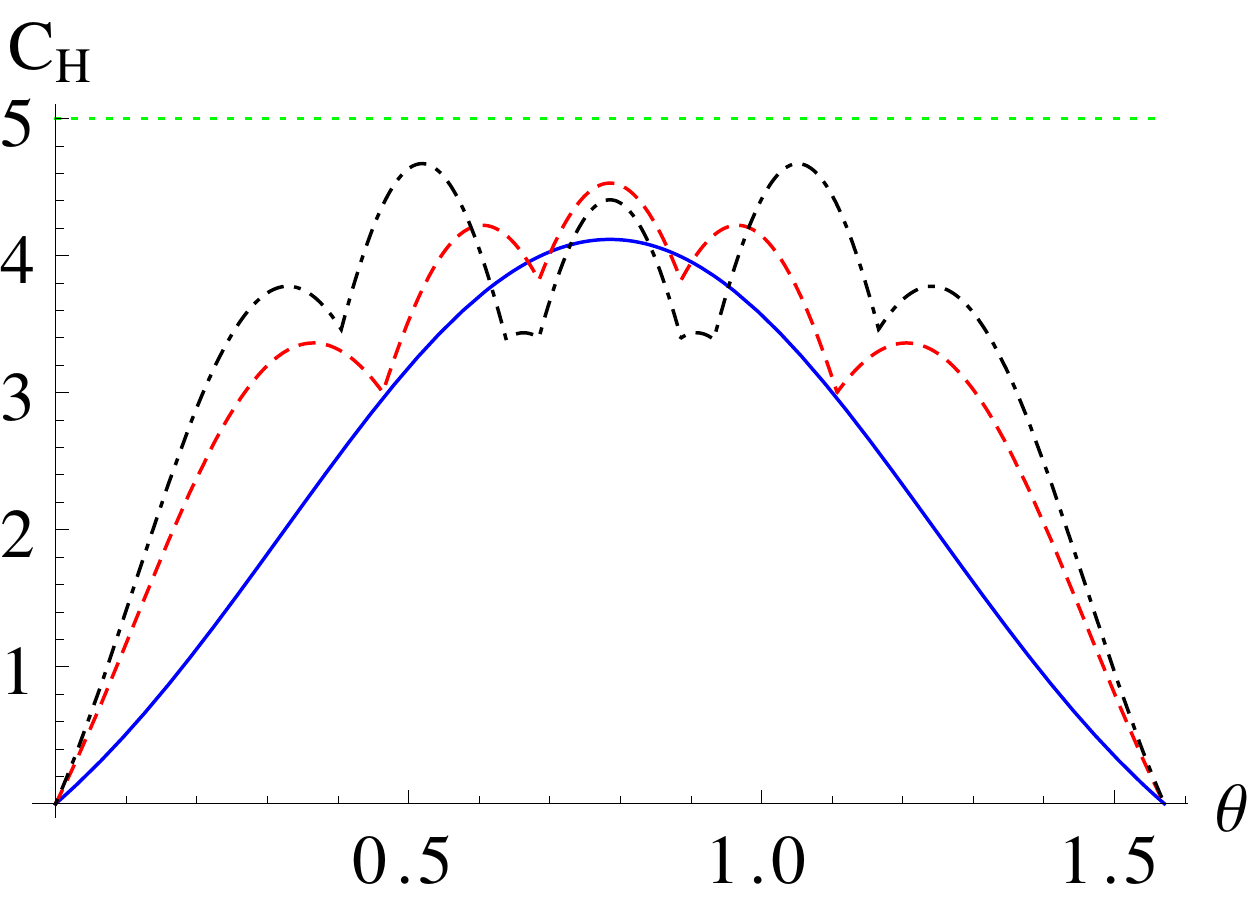}
        \caption{  $C_H$ for the states  $ | {\rm out} \rangle_{5,0}$ (blue solid line), $| {\rm out} \rangle_{4,1}$ (black, dotted-dashed line), and $ | {\rm out} \rangle_{3,2}$ (red, dashed line).  }
        \label{CH5}
    \end{subfigure}\hfill
    \begin{subfigure}[t]{0.32\textwidth}
            \includegraphics[width=\textwidth]{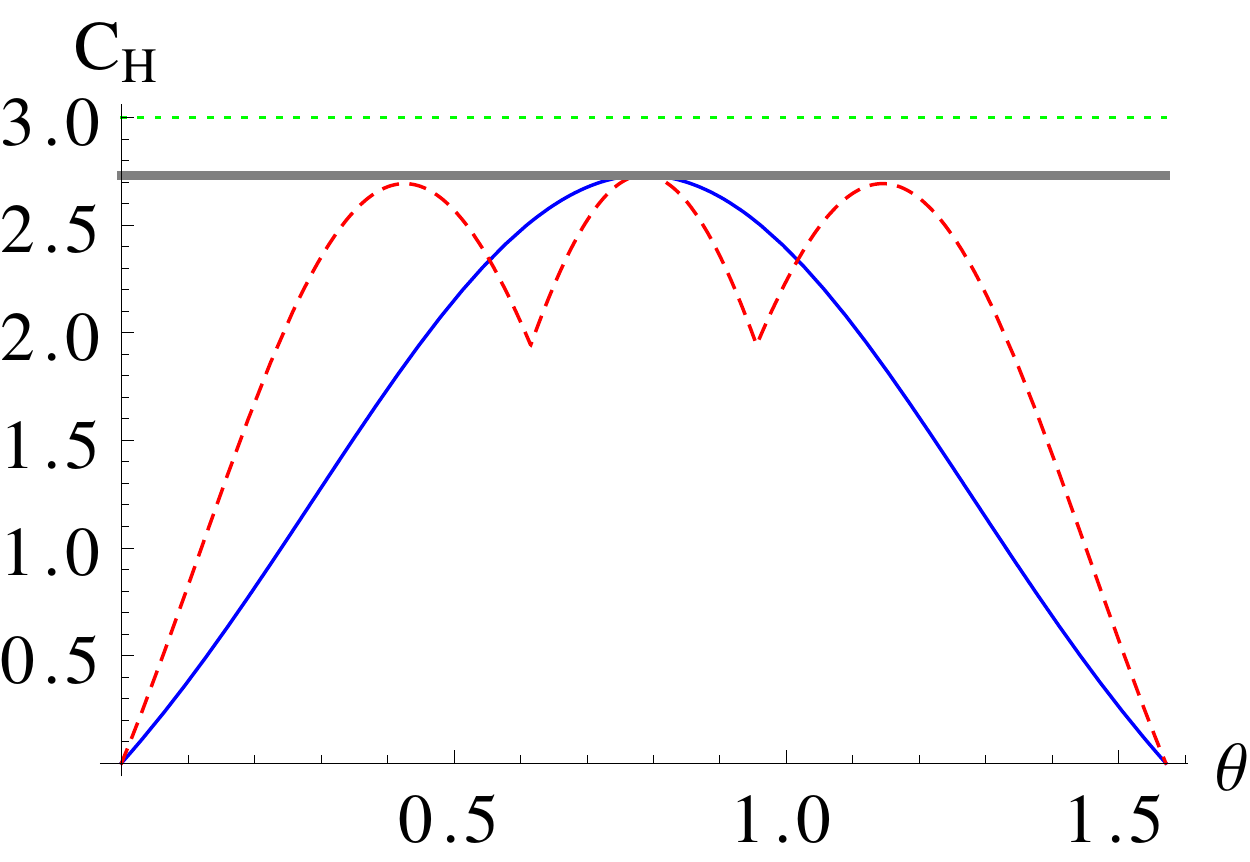}
         \caption{$C_H$ for the states  $ | {\rm out} \rangle_{3,0}$ (blue solid line) and $ | {\rm out} \rangle_{2,1}$ (red, dashed line). } 
        \label{CH3}
    \end{subfigure}\hfill

    \caption{ Plots of the $l1$-norm of coherence for different incoherent input states as functions of the parameter $\theta$. The green, dotted lines mark the maximum coherence in each subspace (\ref{mc}), and the thicker gray line shows the maximum coherence achieved.}
\end{figure*}

For photon number larger than two we present numerical computations confirming the main results already commented. For the case of four photons, displayed in Fig. \ref{CH4}, the larger coherence is obtained by the twin-photon states $ | {\rm out} \rangle_{2,2}$ emerging from an unbalanced beam splitter. Such states $ | {\rm out} \rangle_{2,2}$ are no longer phase-like states, Eq. (\ref{pls}), and so the absolute maximum coherence in Eq. (\ref{mc}) is not reached. Nevertheless, it can be shown numerically that for the optimum case, this is $ | {\rm out} \rangle_{2,2}$ at optimum $\theta$, there is a symmetric split of the photons between the output modes, this is
\begin{equation}
    |\langle n_1,n_2 | {\rm out} \rangle_{2,2} | =     |\langle n_2 ,n_1 | {\rm out} \rangle_{2,2} | .
\end{equation}

\noindent This symmetry feature is specially useful in relation to the sensitivity of two-path interferometers \cite{HFH09}.

\bigskip
Regarding odd number of photons, that is $n_1+n_2 = 2k+1$ for integer $k$, the situation is quite similar, with maximum coherence obtained for input states closer to equal splitting of the photons between input modes, say $|k + 1,k \rangle$, and unbalanced beam splitter $\theta \neq \pi/4$, as shown in Fig. \ref{CH5} for five photons. 

The case of three photons is special, as shown in Fig. \ref{CH3}, since both states $ | {\rm out} \rangle_{3,0}$ (blue solid line) and $ | {\rm out} \rangle_{2,1}$ (red dashed line) reach the maximum coherence for the balanced beam splitter $\theta=\pi/4$, although neither do they achieve the maximum coherence. 

{}

\begin{figure}[h]
    \centering
    \includegraphics[width=8cm]{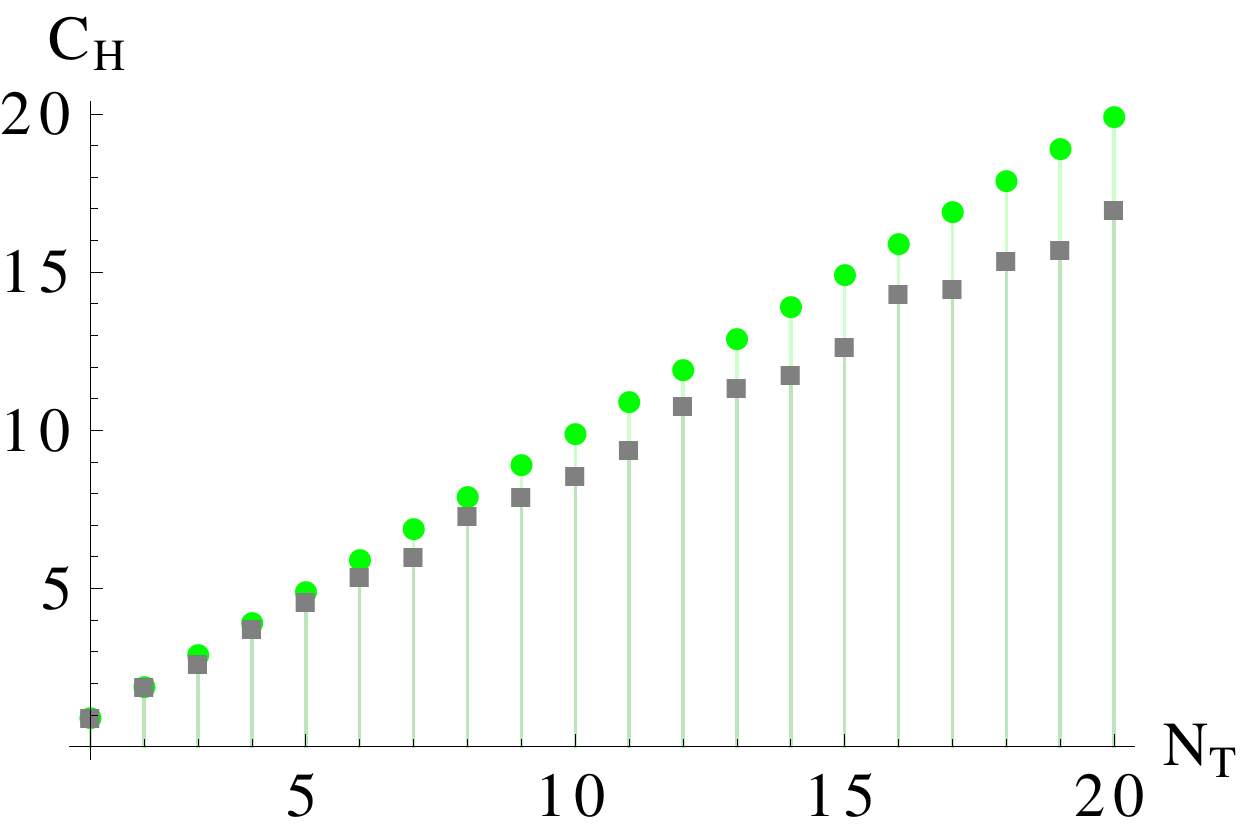}
    \caption{Maximum coherence achieved, $C_{H_{\rm max}}$, for incoming number states on each subspace as a function of the total number of photons involved $N_T$ (gray squares). In green circles, $C_H^{\rm max}=n_1+n_2$.}
    \label{CHBSmaxnumber}
\end{figure}

The maximum coherence scales linearly with the total number of photons but the overlap between the possible outcomes and the phase-states decreases with the total energy so the maximum coherence $C_H^{\rm max}=n_1+n_2$ becomes more and more distant to Eq. (\ref{overlap}). We show this evolution in Fig. \ref{CHBSmaxnumber}.

The results of this section are also reproduced by the relative entropy of coherence.  For the sake of comparison we replicate in Fig. \ref{CHBSmaxnumber} the cases of 3, 4 and 5 photon-subspaces.

\begin{figure*}
\begin{subfigure}{0.32\textwidth}
    \includegraphics[width=\textwidth]{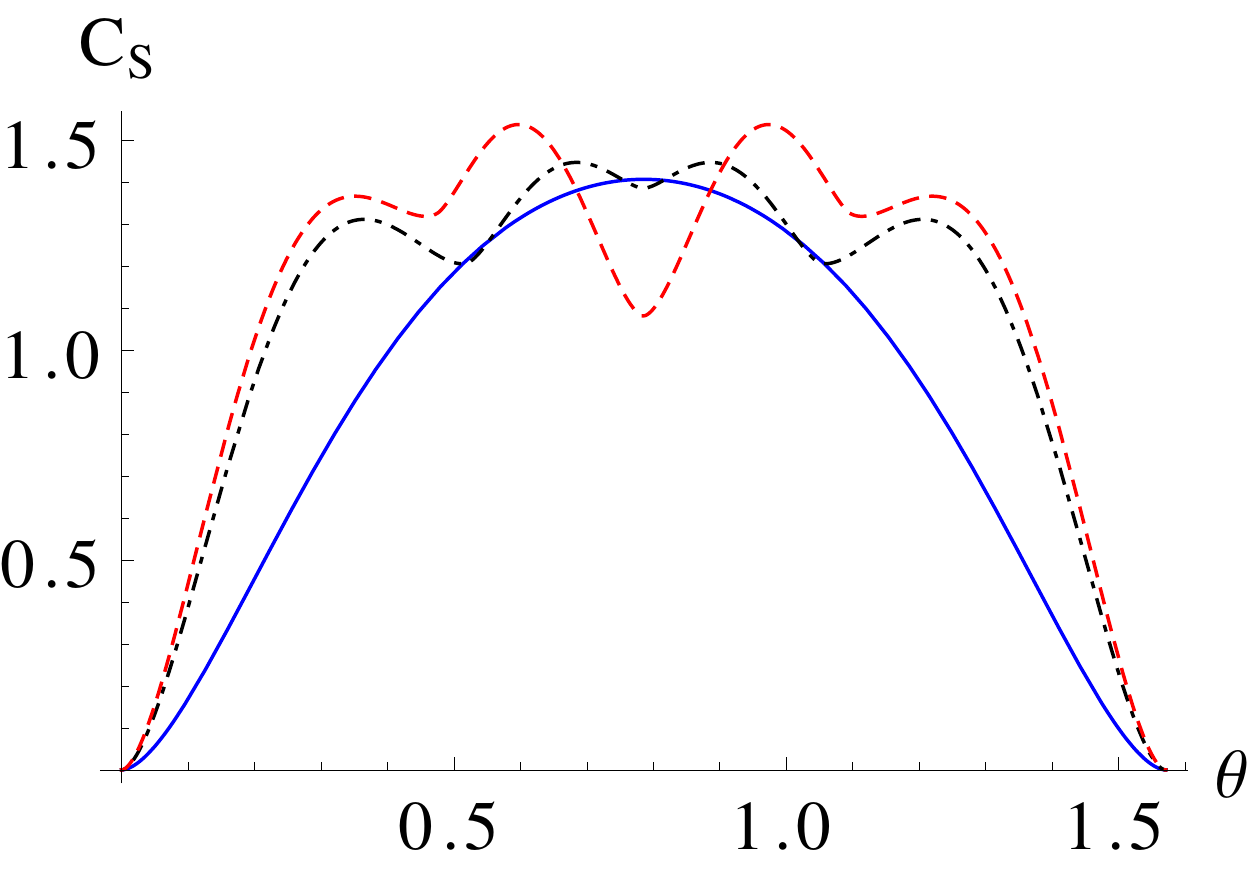}
    \caption{Relative entropy of coherence for the states $|\rm out\rangle_{4,0}$(blue solid line), $| {\rm out} \rangle_{3,1}$ (black, dotted-dashed line), and $ | {\rm out} \rangle_{2,2}$ (red, dashed line), as functions of the parameter $\theta$.}
    \label{CSBSnumber4}
\end{subfigure}\hfill
\begin{subfigure}{0.32\textwidth}
    \includegraphics[width=\textwidth]{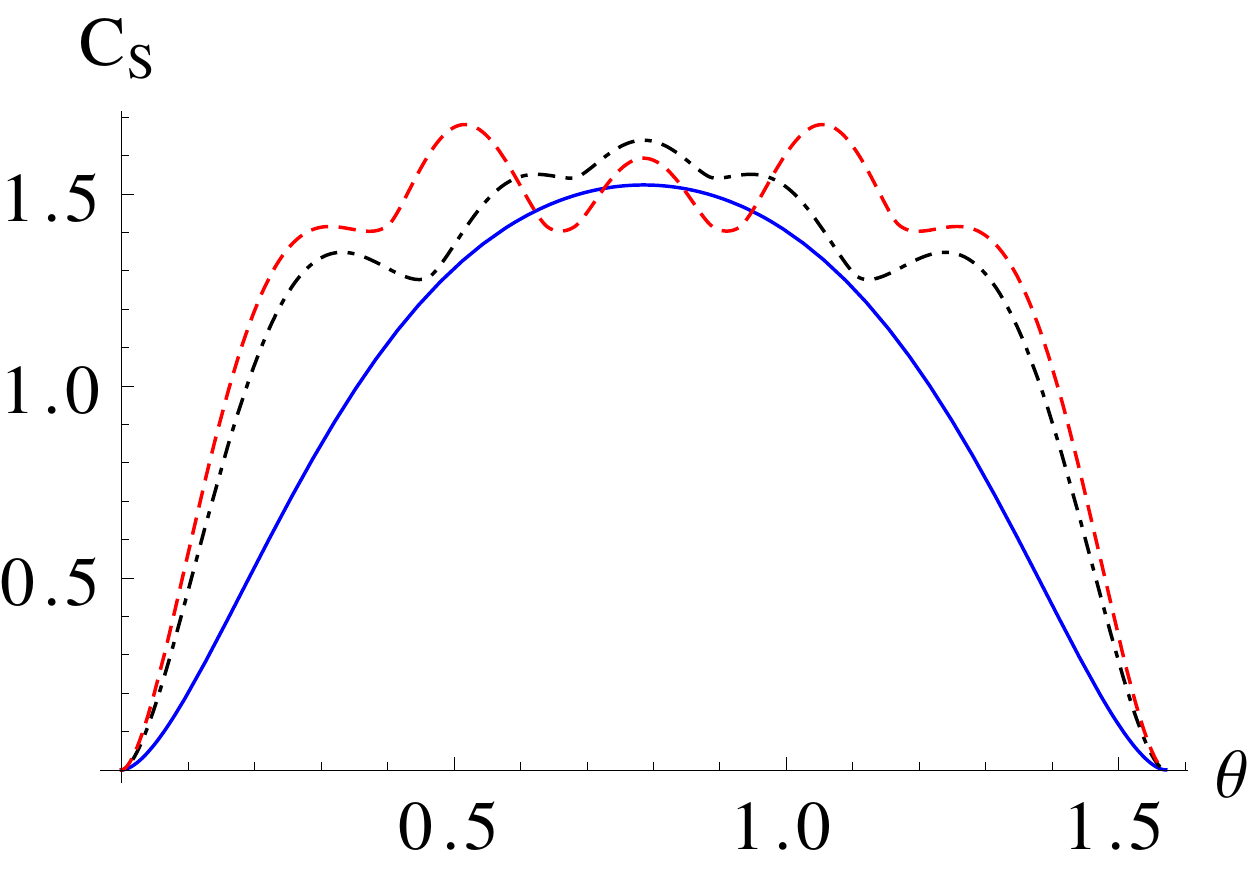}
    \caption{Relative entropy of coherence for the states $|\rm out\rangle_{5,0}$(blue solid line), $| {\rm out} \rangle_{4,1}$ (black, dotted-dashed line), and $ | {\rm out} \rangle_{3,2}$ (red, dashed line), as functions of the parameter $\theta$.}
    \label{CSBSnumber5}
\end{subfigure}\hfill
\begin{subfigure}{0.32\textwidth}
    \includegraphics[width=\textwidth]{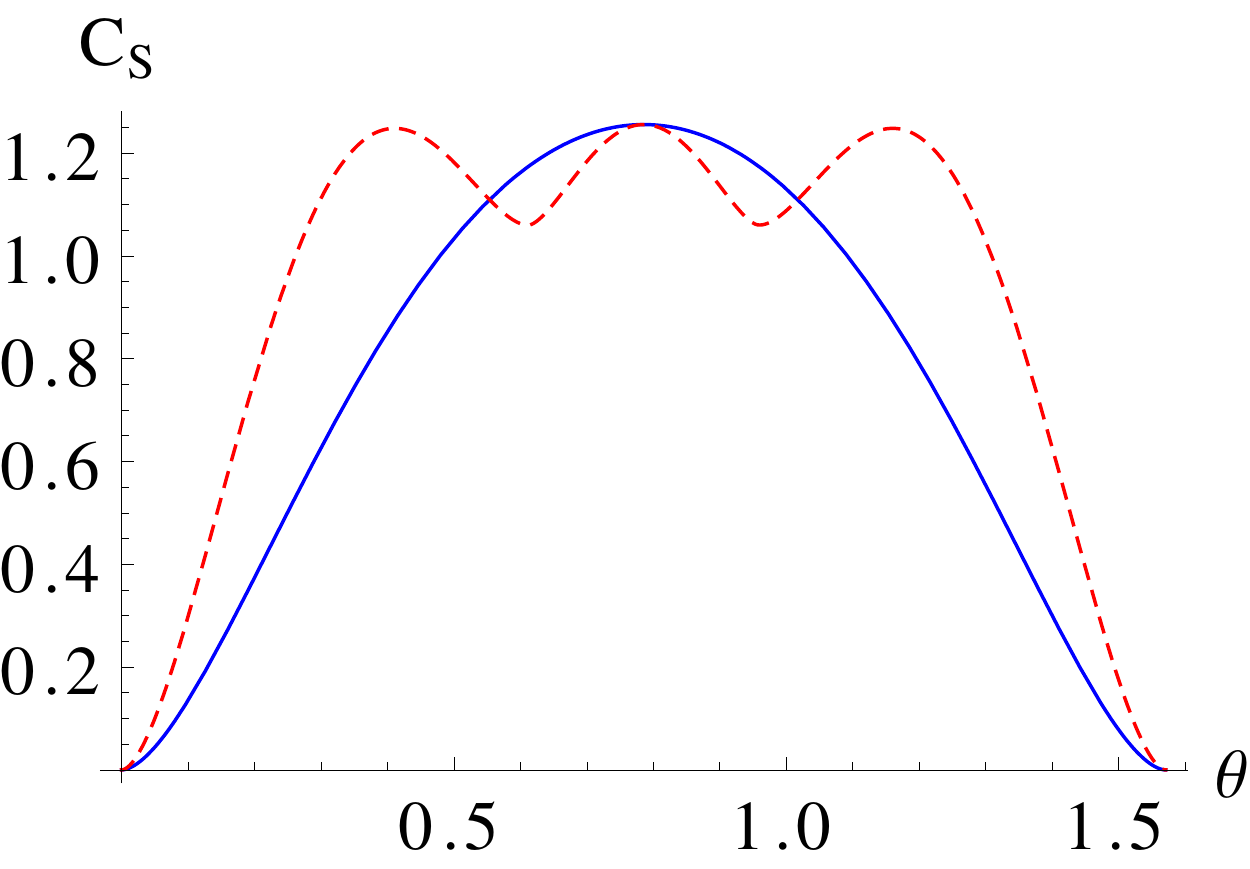}
    \caption{Relative entropy of coherence for the states $|\rm out\rangle_{3,0}$(blue solid line), $| {\rm out} \rangle_{2,1}$  (red,dashed line), as functions of the parameter $\theta$.}
    \label{CSBSnumber3}
    \end{subfigure}
    \label{CSBS}
    \caption{}
\end{figure*}

\begin{figure}[h]
    \centering
    \includegraphics[width=8cm]{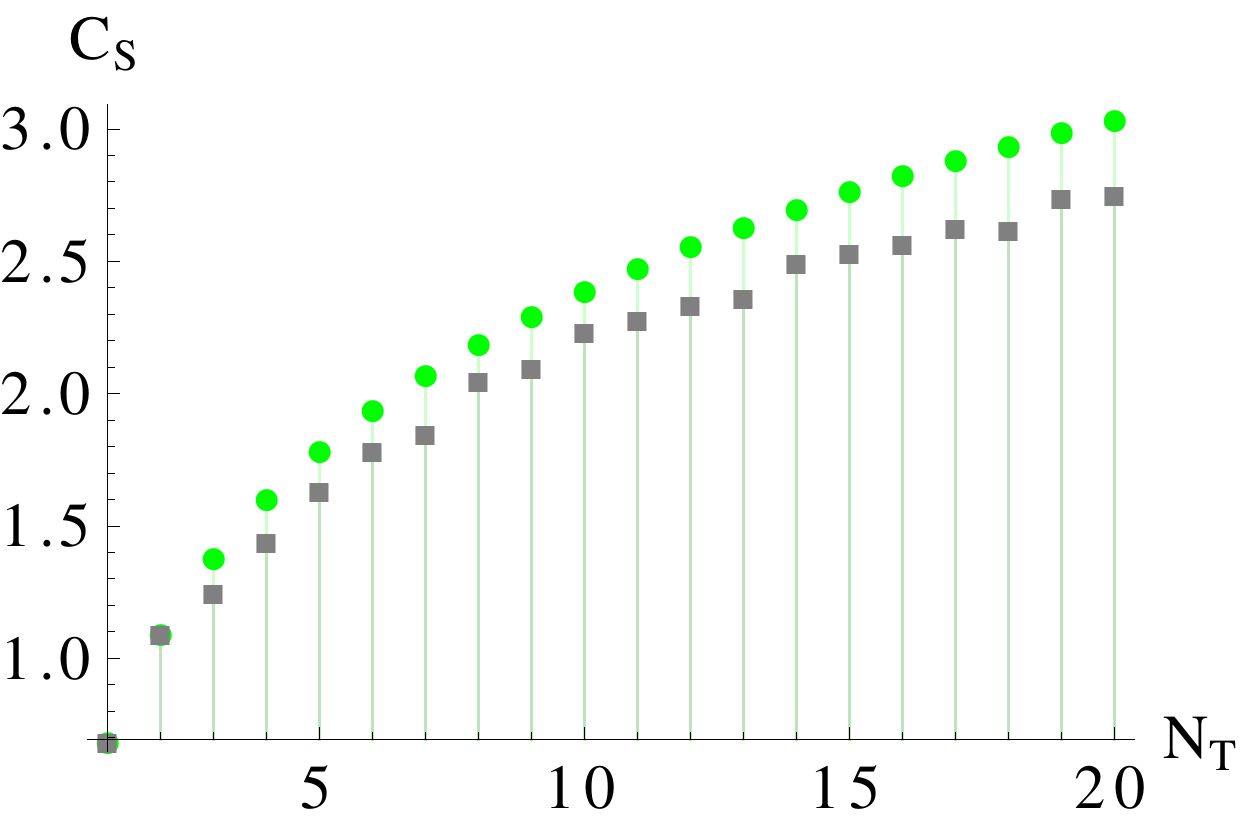}
    \caption{Maximum relative entropy of coherence achieved, $C_{S_{\rm max}}$, for incoming number states on each subspace as a function of the total number of photons involved $N_T$ (gray squares). In green circles, $C_S^{\rm max}=-{\ln}\left[\frac{1}{n_1+n_2+1}\right]$.}
    \label{CSmaxfig}
\end{figure}

Likewise, we can see how the maximum $C_S$ achieved by this incoming states distances from the maximum available coherence in each subspace, Eq. (\ref{CSmax}).
This comparison is presented in Fig. \ref{CSmaxfig}.

\section{Two mode squeezed vacuum}
We take advantage of the previously  computed coherence for the number states to calculate the coherence obtained when each port of the beam splitter is illuminated with one of the modes of the two mode squeezed vacuum state (TMSV),
\begin{equation}
\label{TMSV}
    |\xi\rangle=\sqrt{1-\xi^2}\sum_{n=0}^\infty \xi^n |n,n\rangle,
\end{equation}
where $\xi$ is the squeezing parameter, considered real without loss of generality.
Therefore, we can calculate its coherence as
\begin{equation}
    C_H=\left[\sum_{n=0}^\infty \left(\sqrt{1-\xi^2}  \xi^n \sum_{j=0}^n |c_j| \right)\right]^2-1.
\end{equation}

\begin{figure}
    \centering
    \includegraphics[width=8cm]{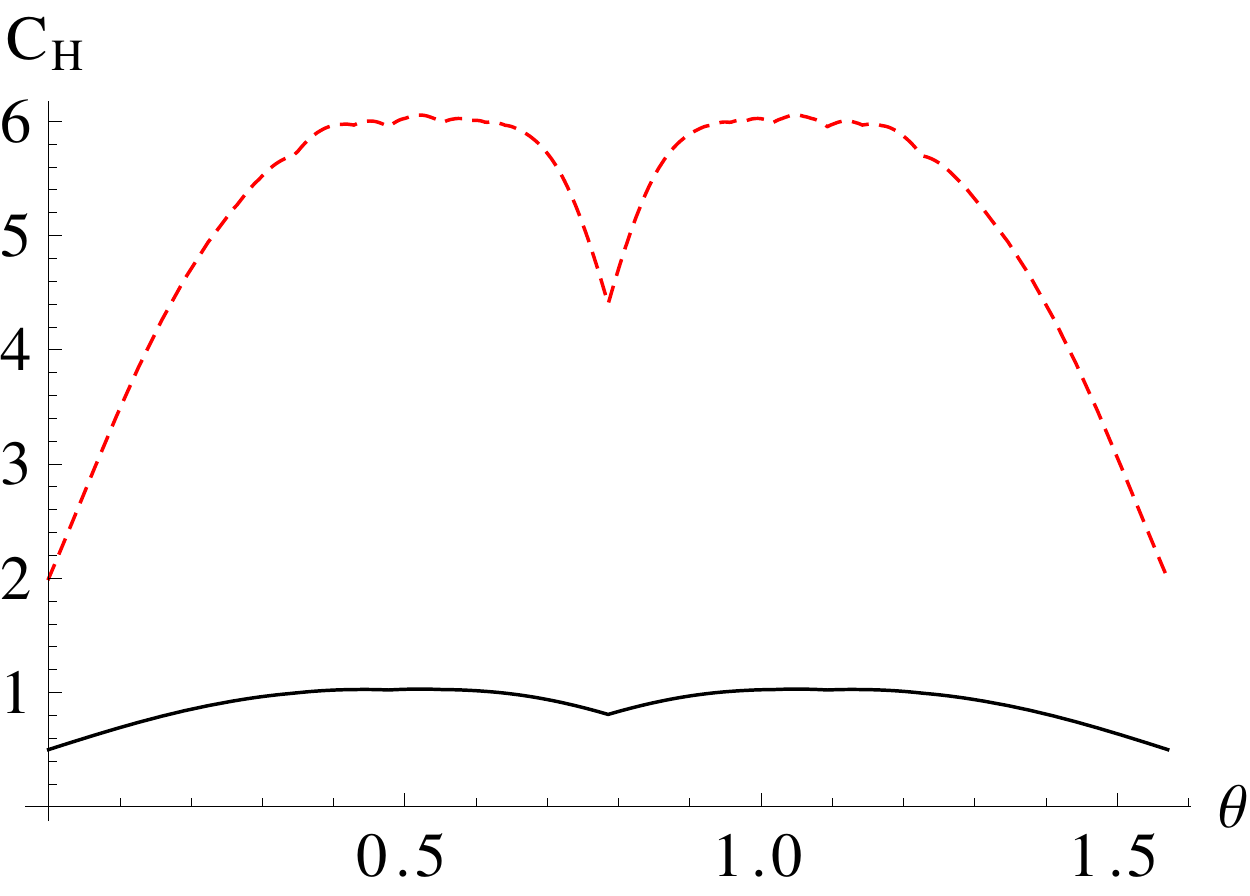}
    \caption{$C_H$ at the output ports of a beam splitter when the inputs are two-mode squeezed vacuum states with $\xi=0.2$ (black, solid line) and $\xi=0.5$ (red, dashed line) as a function of the parameter $\theta$.}
    \label{CHBSTMSV}
\end{figure}

In Fig. \ref{CHBSTMSV} it is represented the final coherence as a function of the parameter $\theta$ for different squeezing parameters $\xi$. The minimum coherence, at $\theta=0$ coincides with the coherence of the squeezed vacuum without any beam splitter transformation \cite{AL22}
\begin{equation}
C_H=\frac{2\xi}{1-\xi}.
\end{equation}

\begin{figure}
    \centering
    \includegraphics[width=8cm]{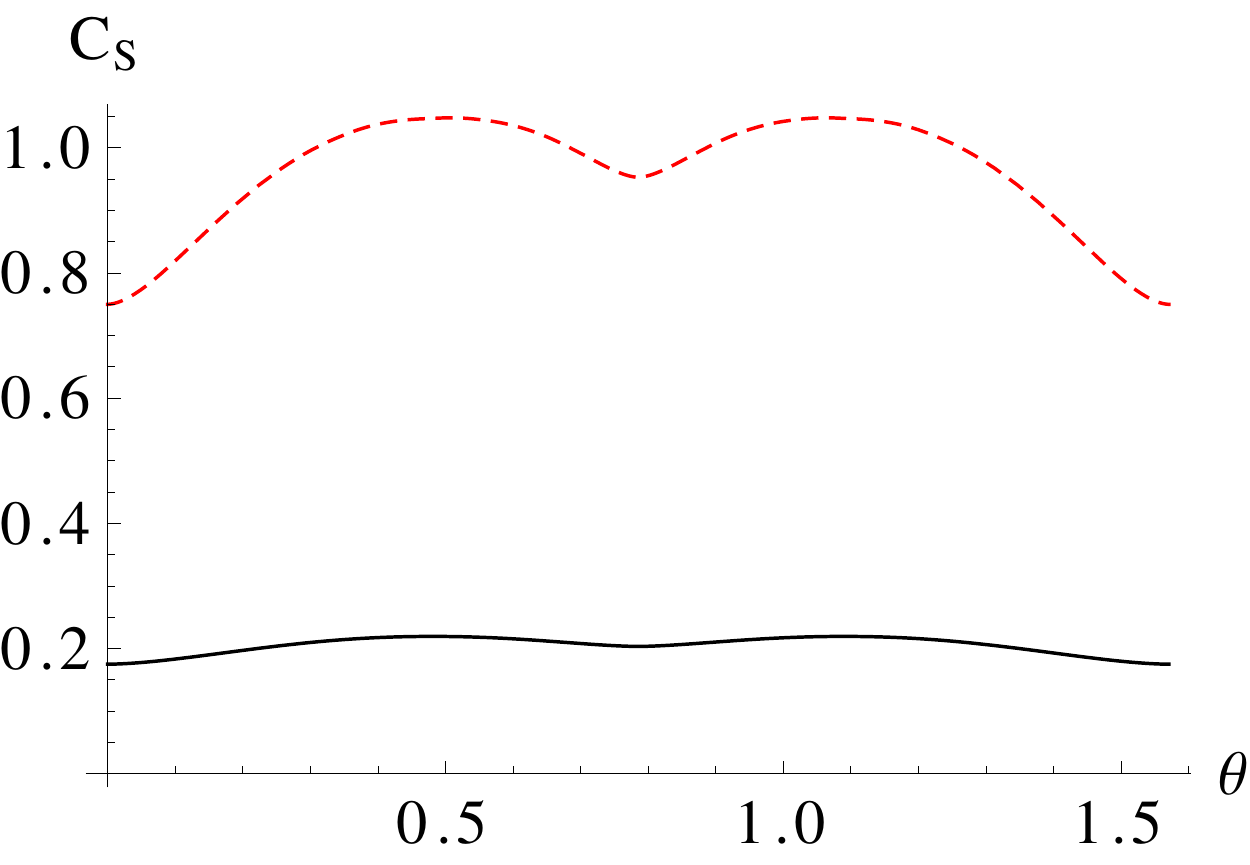}
    \caption{$C_S$ at the output ports of a beam splitter when the inputs are two-mode squeezed vacuum states with $\xi=0.2$ (black, solid line) and $\xi=0.5$ (red, dashed line) as a function of the parameter $\theta$.}
    \label{CSBSTMSV}
\end{figure}

\noindent Once more, the results are replicated by the relative entropy of coherence (see Fig. \ref{CSBSTMSV}).

Since the coherence of the initial state is not zero, we can define the  coherence gained when it goes  through the beam splitter as a function of the incoming amount of coherence. To this end we define the gain in coherence as the next percentage 

\begin{equation}
   \mathcal{G}=\frac{C_{H_{\rm max}}}{C_{H_{\rm min}}}\times 100.
   \label{gain}
\end{equation}

\begin{figure}[h]
    \centering
    \includegraphics[width=8cm]{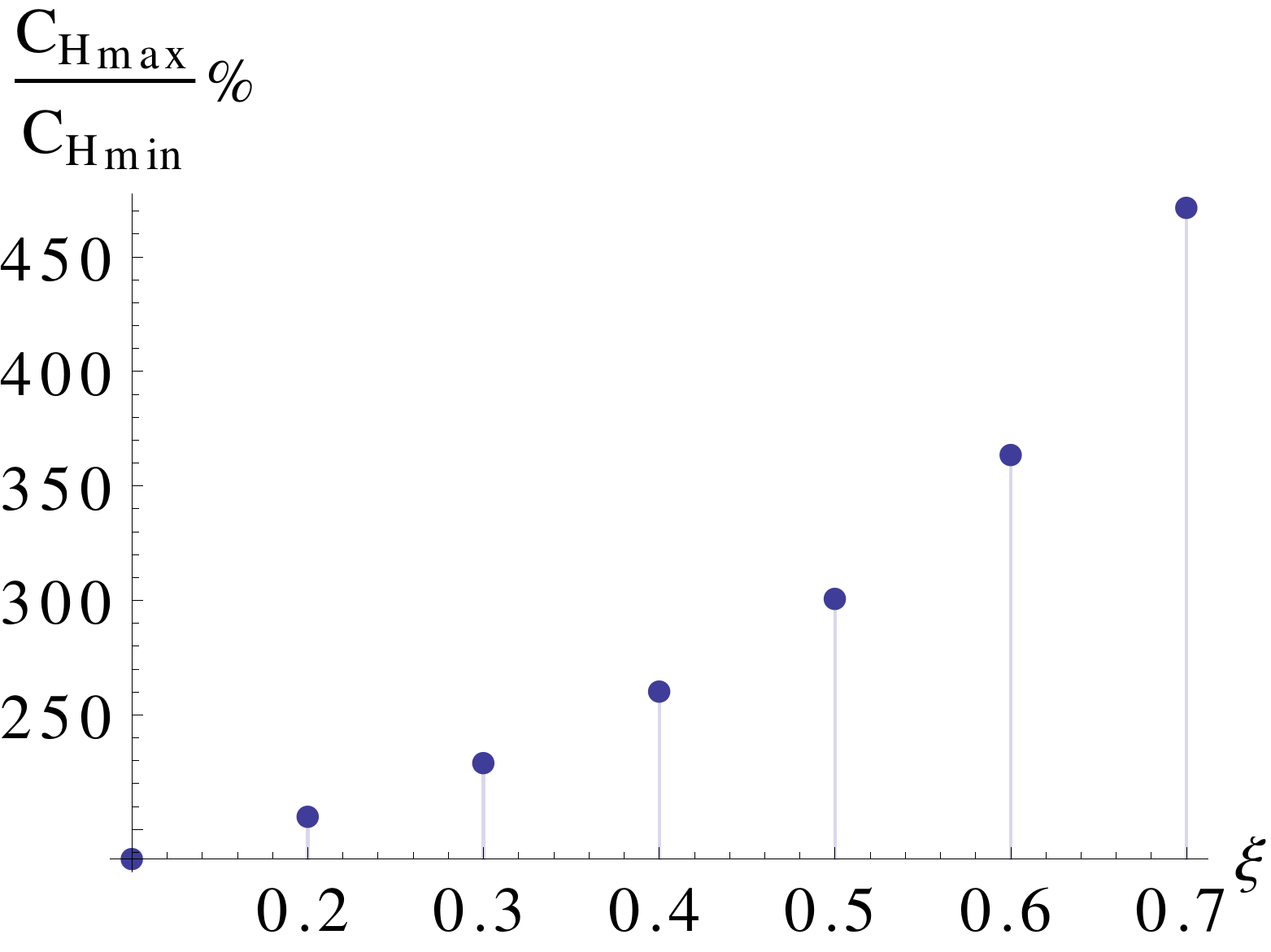}
    \caption{Percentage of coherence enhancement for the two-mode squeezed vacuum as a function of the squeezing parameter $\xi$.}
    \label{CHBSgainTMSV}
\end{figure}

\noindent Different definitions of this concept have been developed \cite{MBP16}. We can compute $\mathcal{G}$ depending on the squeezing of the incoming state, and thus on the mean number of photons. In Fig. \ref{CHBSgainTMSV} it is shown how this gain grows rapidly with the squeezing parameter.

\bigskip

\section{Coherent state}
We compute the coherence provided by the beam splitter when it is illuminated with coherent states on each port. 

We consider the same beam splitter introduced in Eqs. (\ref{io})-(\ref{pr}). The input state in terms of the vacuum state is
\begin{equation}
    \label{coherenta}
    |\alpha_1,\alpha_2\rangle=e^{\alpha_1 a_1^\dagger-\alpha_1^* a_1}e^{\alpha_2 a_2^\dagger-\alpha_2^* a_2}|0,0\rangle,
\end{equation}
and the output becomes

\begin{eqnarray}
    \label{coherentb}
&|{\rm out}\rangle_{\alpha_1\alpha_2}=\\
&|e^{i \tau_1}\alpha_1 \cos{\theta}+e^{i \rho_1}\alpha_2 \sin{\theta},e^{i \tau_2}\alpha_2 \cos{\theta}+e^{i \rho_2}\alpha_1 \sin{\theta}\rangle.\nonumber
\end{eqnarray}

\begin{figure}[h]
    \includegraphics[width=8cm]{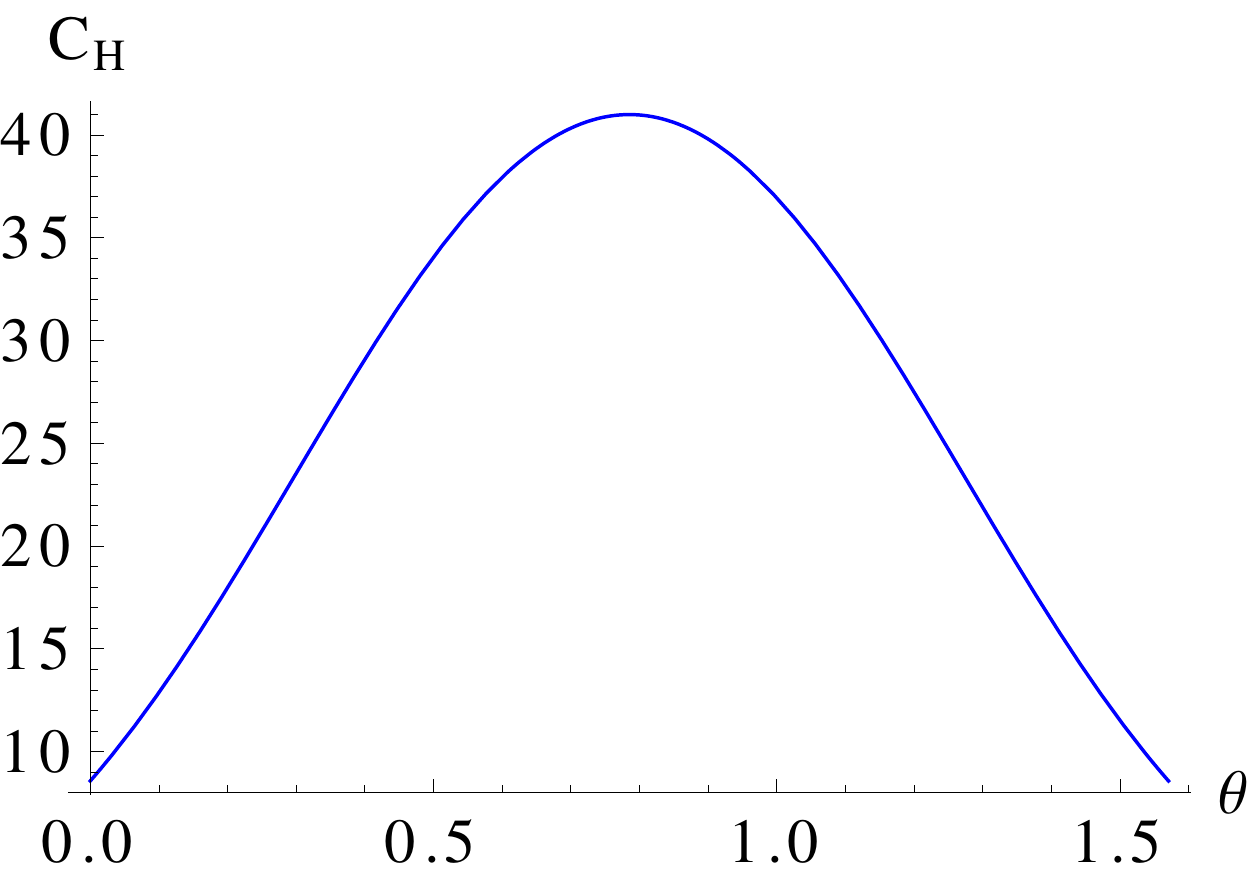}
    \caption{ $\mathcal{C}_H$  for the state  $ | {\rm out} \rangle_{\alpha_1=2,\alpha_2=0}$  as a function of the parameter $\theta$.   }
    \label{CHcohe}
\end{figure}

We compute the coherence of the outcome state when the input state is a coherent coherent of $\bar{N}=4$ on one input port and the vacuum state on the other. In Fig. \ref{CHcohe} it can be seen how the optimum beam splitter configuration is the one that allows a symmetrical output, with the same mean number of photons on each mode.

{}

\bigskip

\begin{figure}[h]
    \includegraphics[width=8cm]{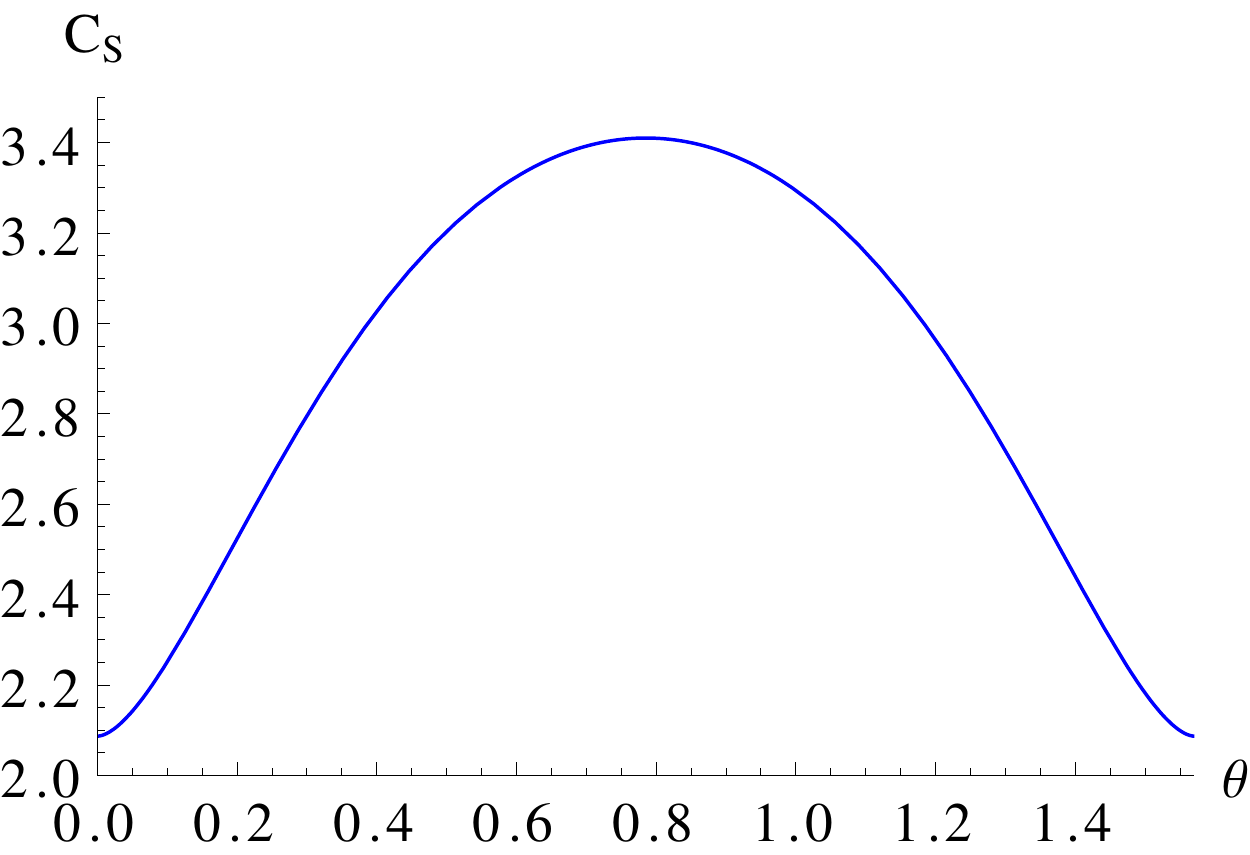}
    \caption{$\mathcal{C}_S$  for the states  $ | {\rm out} \rangle_{\alpha_1=2,\alpha_2=0}$ as functions of the parameter $\theta$.   }
    \label{CScohePlot}
\end{figure}

It can be seen that the minimum of $C_H$ is the coherence of the single-mode coherent state $C_{H_{min}}=C_H(|\alpha=\sqrt{\bar{N}}\rangle)$ and it appears when the output state is of the form $|\alpha=\sqrt{\bar{N}},0\rangle$.
As in the incoherent case, this result can be reproduced by the relative entropy of coherence as illustrated in Fig. \ref{CScohePlot}.
{}

\bigskip

\begin{figure}[h]
    \includegraphics[width=8cm]{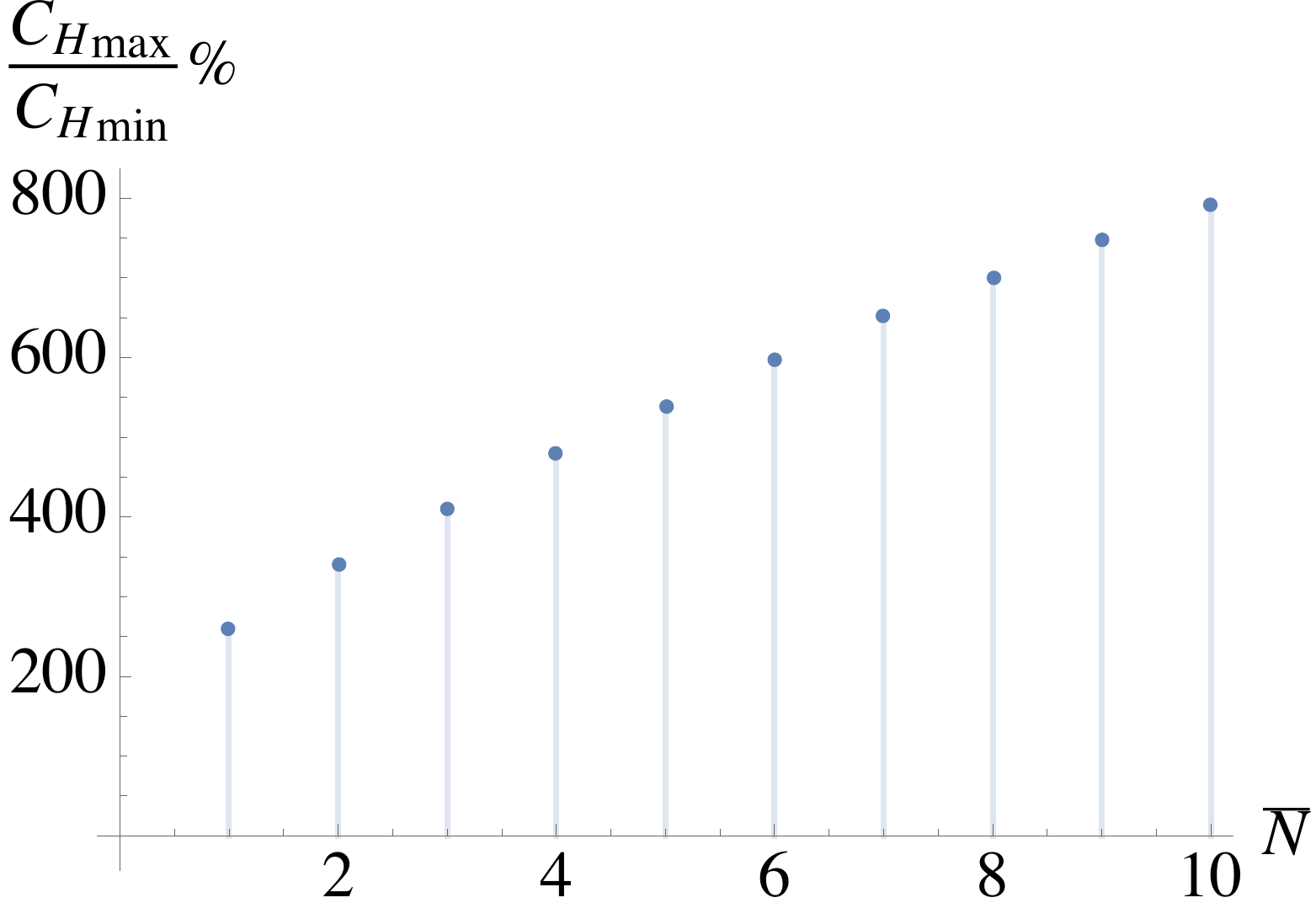}
    \caption{ Plot of the percentage of coherence gained as a function of the mean number of photons.}
    \label{CHcohegain}
\end{figure}

We may analyze the advantage of the beam splitter in terms of coherence by computing the gain in Eq. (\ref{gain}). In Fig. \ref{CHcohegain} it can be seen how the gain of coherence increases with the initial coherence of the single-mode state but the growth ratio is smaller for high $\bar{N}$.

{}

\subsection{Comparison with TMSV}

In the following we compare the performance of the beam splitter when illuminated by the coherent state and by the TMSV when these present identical: a) minimum amount of coherence, $C_{H_{min}}$, and b) mean number of photons, $\bar{N}$.

As introduced, we are considering two states with the same minimum amount of coherence, e. g.  $C_{H_{min}}=3.0$. The corresponding coherent state before the beam splitter is $|\alpha_1=\sqrt{0.83},0\rangle$, and the gain in coherence caused 
by the transformation is

\begin{equation}
\mathcal{G}_\alpha\approx245\%. 
\end{equation}

 However, if we consider a TMSV  with the same minimum coherence available, this is $|\xi=0.6\rangle$, the gain produced by the beam splitter is remarkably higher, 
\begin{equation}
\mathcal{G}_\xi\approx364\%.
\end{equation}

\bigskip

Regarding the energy, the mean number of photons of the TMSV is
\begin{equation}
    \bar N=\frac{\xi^2}{1-\xi^2},
\end{equation}
thus, for $\bar N=1$ the gain in coherence caused by the beam splitter is 
\begin{equation}
\mathcal{G}_\xi\approx470\%
\end{equation}
whereas for a coherent state with $|\alpha|^2=1$ it is \begin{equation}
\mathcal{G}_\alpha\approx260\%.
\end{equation}

Therefore, the TMSV is able to gain more coherence since the squeezing allows it to better resemble the constant statistics of phase-like states.

\section{Conclusions}
We have performed a detailed study of the role of beam splitters as quantum coherence makers, obtaining the optimum configuration of the reflectance-transmitance parameters for several incoming states. The optimum configuration is such that the outcoming state is as similar as possible to the phase-like states. We have investigated how the amount of energy of the input state concerns the maximum coherence achievable by the system. By studying, at a fixed energy, the two mode squeezed vacuum along with the coherent state cases we conclude that the beam splitter generates considerably more coherence when illuminated by the TMSV. In the same way, the  gain in coherence of the TMSV is higher if we consider fixed the minimum coherence available.

\noindent{\bf Acknowledgments.- }
L. A. acknowledges financial support from the Spanish Ministerio de Ciencia e Innovaci\'on and the European Social Fund, Contract Grant No. BES-2017-081942.

\end{document}